\documentclass[aps,prl,twocolumn,showpacs,groupedaddress]{revtex4}

\usepackage{graphicx}
\usepackage{dcolumn}
\usepackage{bm}
\usepackage{amssymb}

\usepackage[usenames]{color}

\begin{document}

\newcommand{\dzero}     {D0}
\newcommand{\met}       {\mbox{$\not\!\!E_T$}}
\newcommand{\deta}      {\mbox{$\eta^{\rm det}$}}
\newcommand{\meta}      {\mbox{$\left|\eta\right|$}}
\newcommand{\mdeta}     {\mbox{$\left|\eta^{\rm det}\right|$}}
\newcommand{\rar}       {\rightarrow}
\newcommand{\rargap}    {\mbox{ $\rightarrow$ }}
\newcommand{\tbbar}     {\mbox{$tb$}}
\newcommand{\tqbbar}    {\mbox{$tqb$}}
\newcommand{\ttbar}     {\mbox{$t\bar{t}$}}
\newcommand{\bbbar}     {\mbox{$b\bar{b}$}}
\newcommand{\ccbar}     {\mbox{$c\bar{c}$}}
\newcommand{\qqbar}     {\mbox{$q\bar{q}$}}
\newcommand{\ppbar}     {\mbox{$p\bar{p}$}}
\newcommand{\comphep}   {\sc{c}\rm{omp}\sc{hep}}
\newcommand{\herwig}    {\sc{herwig}}
\newcommand{\pythia}    {\sc{pythia}}
\newcommand{\alpgen}    {\sc{alpgen}}
\newcommand{\singletop} {\rm{SingleTop}}
\newcommand{\reco}      {\sc{reco}}
\newcommand{\Mchiggs}   {\mbox{$M({\rm jet1,jet2},W)$}}
\newcommand{\coss}	{\mbox{\rm{cos}$\theta^{\star}$}}
\newcommand{\ljets} {$\ell+$jets}

\lefthyphenmin=6
\righthyphenmin=6

\hspace{5.2in}\mbox{Fermilab-Pub-08/583-E}

\title{Search for anomalous top quark couplings with the D0 detector} 
%
\author{V.M.~Abazov$^{36}$}
\author{B.~Abbott$^{75}$}
\author{M.~Abolins$^{65}$}
\author{B.S.~Acharya$^{29}$}
\author{M.~Adams$^{51}$}
\author{T.~Adams$^{49}$}
\author{E.~Aguilo$^{6}$}
\author{M.~Ahsan$^{59}$}
\author{G.D.~Alexeev$^{36}$}
\author{G.~Alkhazov$^{40}$}
% ...

\affiliation{\vspace{0.1 in}(The D\O\ Collaboration)\vspace{0.1 in}}
% ...
\affiliation{$^{6}$University of Alberta, Edmonton, Alberta, Canada,
                Simon Fraser University, Burnaby, British Columbia, Canada,
                York University, Toronto, Ontario, Canada, and
                McGill University, Montreal, Quebec, Canada}
% ...
\affiliation{$^{29}$Tata Institute of Fundamental Research, Mumbai, India}
% ...
\affiliation{$^{36}$Joint Institute for Nuclear Research, Dubna, Russia}
% ...
\affiliation{$^{40}$Petersburg Nuclear Physics Institute,
                St. Petersburg, Russia}
% ...
\affiliation{$^{49}$Florida State University, Tallahassee, Florida 32306, USA}
% ...
\affiliation{$^{51}$University of Illinois at Chicago,
                Chicago, Illinois 60607, USA}
% ...
\affiliation{$^{59}$Kansas State University, Manhattan, Kansas 66506, USA}
% ...
\affiliation{$^{65}$Michigan State University,
                East Lansing, Michigan 48824, USA}
% ...
\affiliation{$^{75}$University of Oklahoma, Norman, Oklahoma 73019, USA}
% ...

\date{December 31, 2008}

\begin{abstract}
Anomalous $Wtb$ couplings modify the angular correlations of the top quark decay
products and change the single top quark production cross section. We present limits on 
anomalous top quark couplings by combining information from $W$~boson helicity measurements in top 
quark decays and anomalous coupling searches in the single top quark final state. 
We set limits on right-handed vector couplings as 
well as left-handed and right-handed tensor couplings based on about 1~fb$^{-1}$ of data collected by 
the D0 experiment.


\end{abstract}
\maketitle

\vspace{-0.1in}
The top quark is by far the heaviest fermion in the standard model (SM), and thus has the strongest coupling
to the Higgs boson of all SM fermions. 
This makes the top quark and its interactions an ideal place to look for new physics related to 
electroweak symmetry breaking. The coupling between the top quark and the weak gauge bosons may be 
altered by physics beyond the SM. In particular the coupling between the top quark and the $W$~boson
determines most of the top quark phenomenology and can be sensitively probed at 
hadron colliders~\cite{TT_window}. The effective Lagrangian describing the $Wtb$ 
interaction including operators up to dimension five is~\cite{cpyuan_0503040v3}:
\begin{eqnarray}
\mathcal{L}&=&-\frac{g}{\sqrt{2}}\bar{b} \gamma^\mu V_{tb}
\nonumber (f^L_1 P_L + f^R_1 P_R) t W_{\mu}^{-}\\
&-&\frac{g}{\sqrt{2}} \bar{b} \frac{i\sigma^{\mu\nu} q_{\nu}V_{tb}}{M_W} (f^L_2 P_L + f^R_2 P_R) t W_{\mu}^{-} 
+ h.c. \, ,
\label{coupling}
\end{eqnarray}
 where  $M_W$ is the mass of the $W$~boson, $q_{\nu}$ is its four-momentum, $V_{tb}$ is the 
Cabibbo-Kobayashi-Maskawa matrix element~\cite{Cabibbo:1963yz}, and $P_{L}=(1 - \gamma_5)/2$ 
($P_{R}=(1 + \gamma_5)/2$) is the left-handed (right-handed) projection operator.  
In the SM, the $Wtb$ coupling is purely left-handed, and the values
of the coupling form factors are $f^{L}_{1} \approx 1 $, ${f^{L}_{2}=f^{R}_{1}=f^{R}_{2}=0}$. We assume real 
coupling form factors, implying $CP$ conservation, and a spin-$\frac{1}{2}$ top quark which decays
predominantly to $Wb$.
Indirect constraints on the magnitude of the right-handed vector coupling and tensor couplings 
exist from measurements of the $b \rightarrow s\gamma$ branching fraction~\cite{bsgamma}. While those limits
are tighter than the direct limits presented here, they also include assumptions that are 
not required here, for example the absence of other sources of new physics coupling to the $b$~quark.

We search for non-SM values of the couplings using $\approx~$1 fb$^{-1}$ of data collected by the 
D0 experiment~\cite{NIM} at the Fermilab Tevatron $p\bar{p}$ collider between 2002 and 2006 (Run~II).  
Variations in the coupling form factors would mainly manifest themselves in two distinct ways at D0:  
by changing the rate and kinematic distributions of electroweak single top quark production, 
and by altering the fractions of $W$~bosons from top quark decays produced in each of the three possible 
helicity states.  In this Letter, we combine information from our measurement of the $W$~boson helicity 
fractions in $t\bar{t}$ events~\cite{Whel-d0} with information from single top quark production.
We have previously set direct limits on anomalous top quark coupling form factors based solely on the
single top quark final state~\cite{wtb-prl}. Here we set substantially tighter limits
on the effective top quark couplings using the general framework given in Ref.~\cite{Chen:2005vr}.
This is the first such combination of all applicable D0 Run~II top quark measurements to limit anomalous
top quark coupling form factors.

We follow the approach adopted in Ref.~\cite{wtb-prl} and investigate one pair of coupling form factors at a time
out of the full set of form factors ($f^{L}_{1}$, $f^{R}_{1}$, $f^{L}_{2}$, and 
$f^{R}_{2}$. For each pair under investigation we assume that the other two have the SM values. 
We consider three cases, pairing the left-handed vector coupling form factor $f^{L}_{1}$ with
each of the other three form factors. We refer to these as $(L_1,R_1)$, $(L_1,L_2)$, and $(L_1,R_2)$.
For each pair of form factors a likelihood distribution is extracted from the $W$~helicity measurement of
the decay angle distribution in top quark decays. All top quark pair events with decays to at least 
one lepton (electron or muon) are included in the $W$~helicity measurement. This likelihood is then 
combined with the result of the anomalous
couplings search in the single top quark final state in a Bayesian statistical analysis, yielding
a two-dimensional posterior probability density as a function of both form factors. We extract
limits on $f^{R}_{1}$, $f^{L}_{2}$, and $f^{R}_{2}$ by projecting the two-dimensional posterior onto 
the corresponding form factor axis.

The $W$ boson helicity measurement, described in Ref.~\cite{Whel-d0}, uses events in both 
the $\ell+$jets ($t\bar{t}\rightarrow W^+W^-b\bar{b}\rightarrow \ell\nu q\bar{q}^{\prime}b\bar{b}$) and 
dilepton ($t\bar{t}\rightarrow W^+W^-b\bar{b}\rightarrow \ell\nu\ell^{\prime}\nu^{\prime}b\bar{b}$)  
final states.  The measurement variable is $\theta^*$, the angle between the down-type fermion and top 
quark momenta in the $W$ boson rest frame.  To evaluate this variable, we assign a momentum to the neutrino(s) 
either via a constrained kinematic fit (in the $\ell+$jets channel) or an algebraic solution (in the 
dilepton channel).

We use the {\alpgen} leading-order Monte Carlo (MC) event generator~\cite{alpgen}, 
interfaced to {\sc pythia}~\cite{pythia}, to model {\ttbar} events as well as $W$+jets and $Z$+jets 
background events. We generate both SM $V-A$ and $V+A$ $Wtb$ couplings, and reweight events to model 
a given $W$ boson helicity state. We use the CTEQ6L1 parton distribution functions~\cite{cteq} and
set the top quark mass to 172.5~GeV. 
The response of the D0 detector to the MC events is simulated 
using {\sc geant}~\cite{geant}. 
We model the background from multijet production where a jet is misidentified as an isolated electron or muon 
using events from data containing lepton candidates which pass all of the lepton identification 
requirements except one but otherwise resemble the signal events.
We use MC to model other small backgrounds (diboson and single top quark production).

We select events with a multivariate likelihood discriminant that uses both kinematic and 
$b$-lifetime information to distinguish $t\bar{t}$ events from background and obtain a
sample of 288 \ljets~(75 dilepton) events with an expected background contribution of 
$54 \pm 7$ ($17 \pm 4$) events.

\begin{figure}[!h!btp]
\vspace{-0.1in}
\hspace{-0.02\textwidth}
\includegraphics[width=0.253\textwidth]{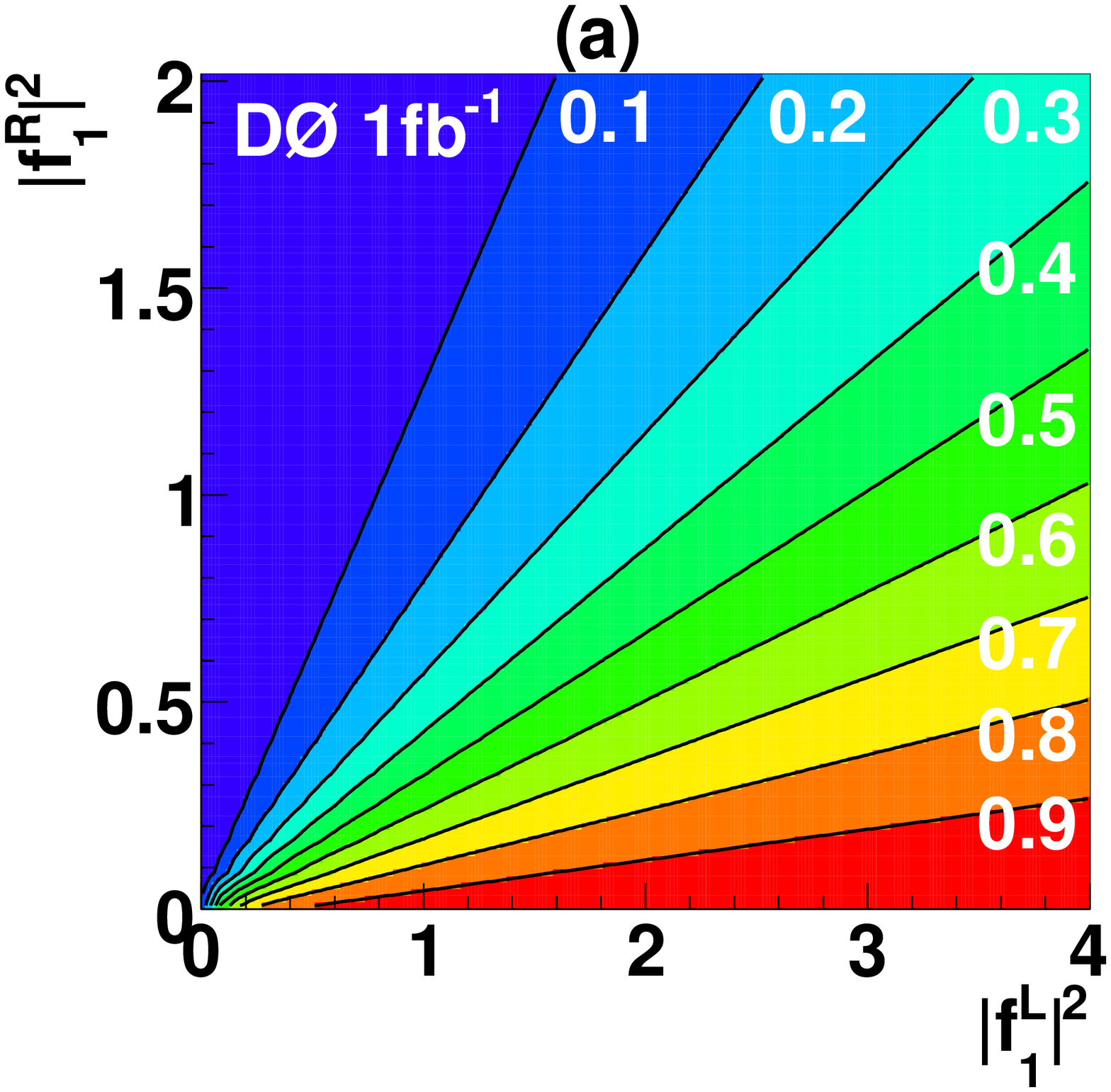} \hspace{-0.02\textwidth}
\vspace{-0.05in}
\includegraphics[width=0.253\textwidth] {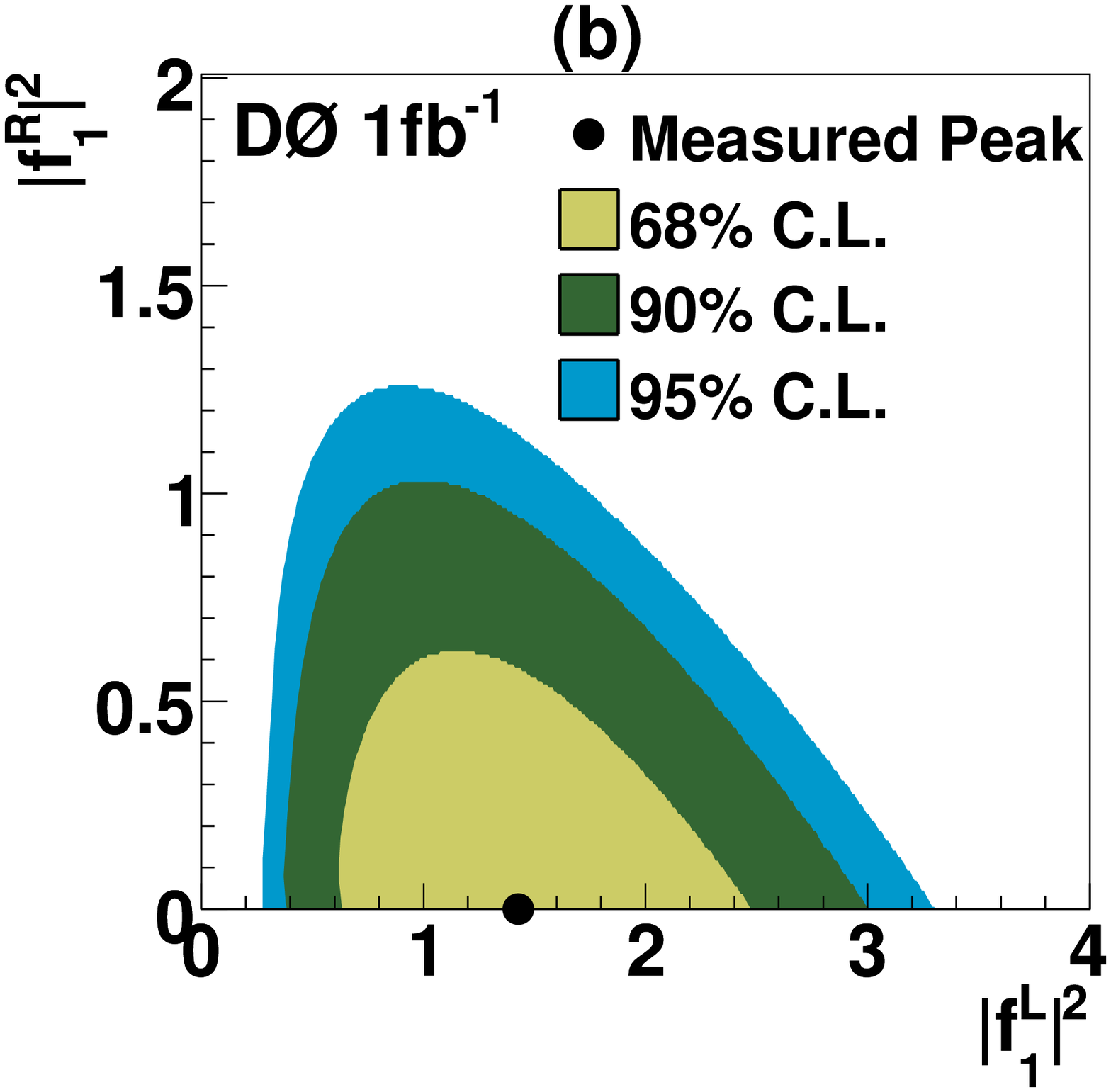}
\vspace{-0.05in}
\hspace{-0.02\textwidth}
\includegraphics[width=0.253\textwidth]{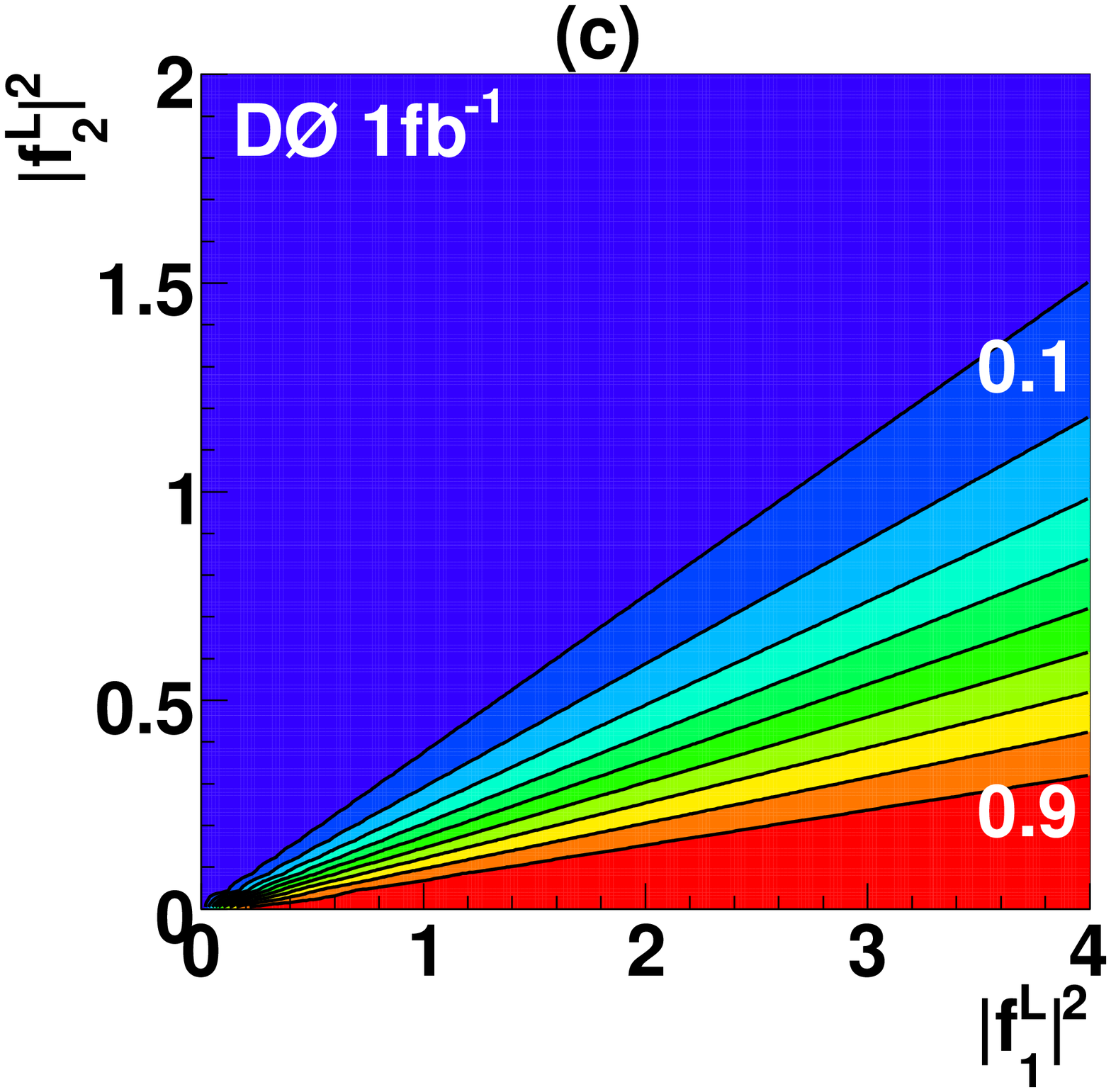} \hspace{-0.02\textwidth}
\includegraphics[width=0.253\textwidth] {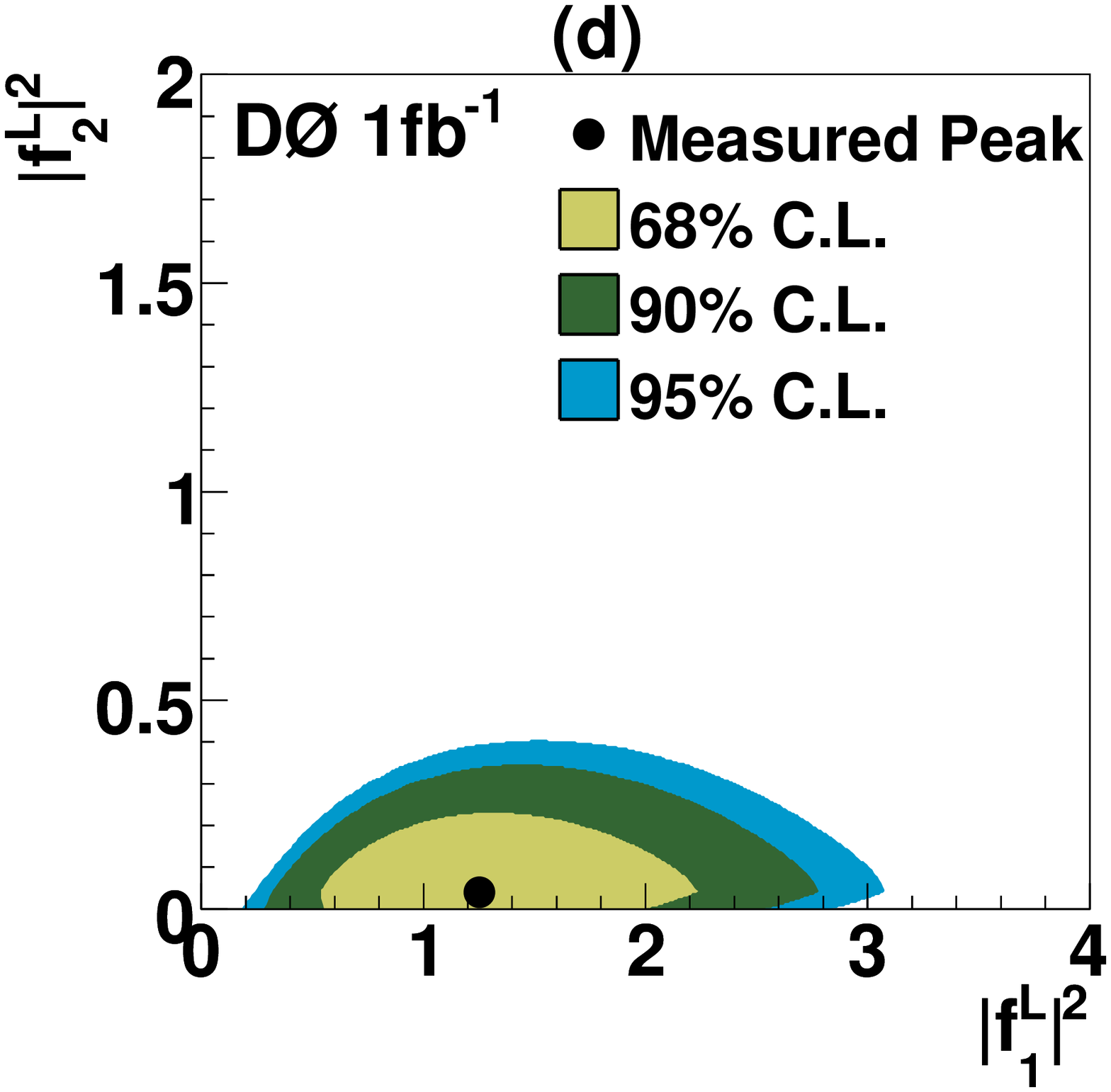}
\vspace{-0.05in}
\hspace{-0.02\textwidth}
\includegraphics[width=0.253\textwidth]{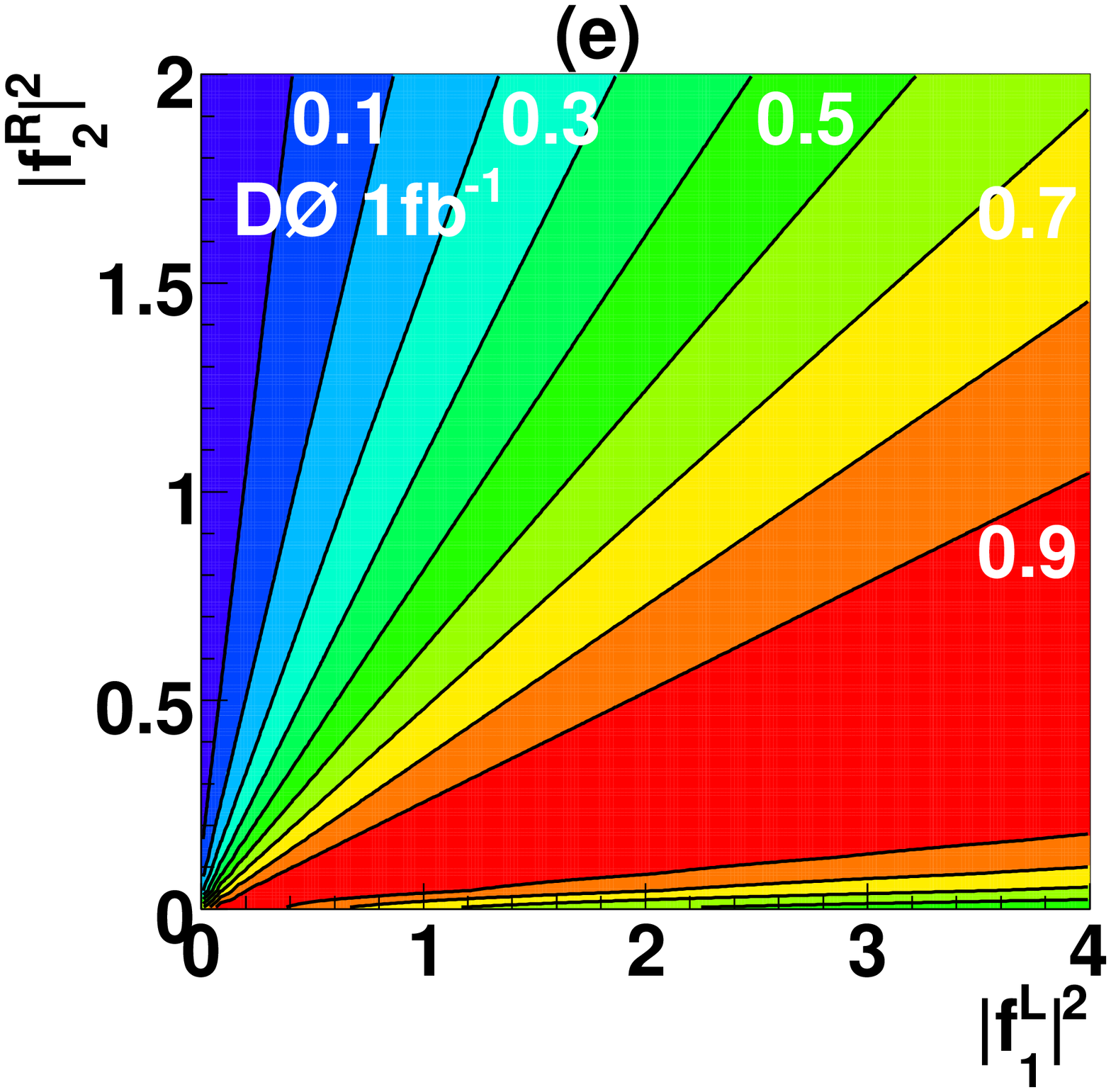} \hspace{-0.022\textwidth}
\includegraphics[width=0.253\textwidth] {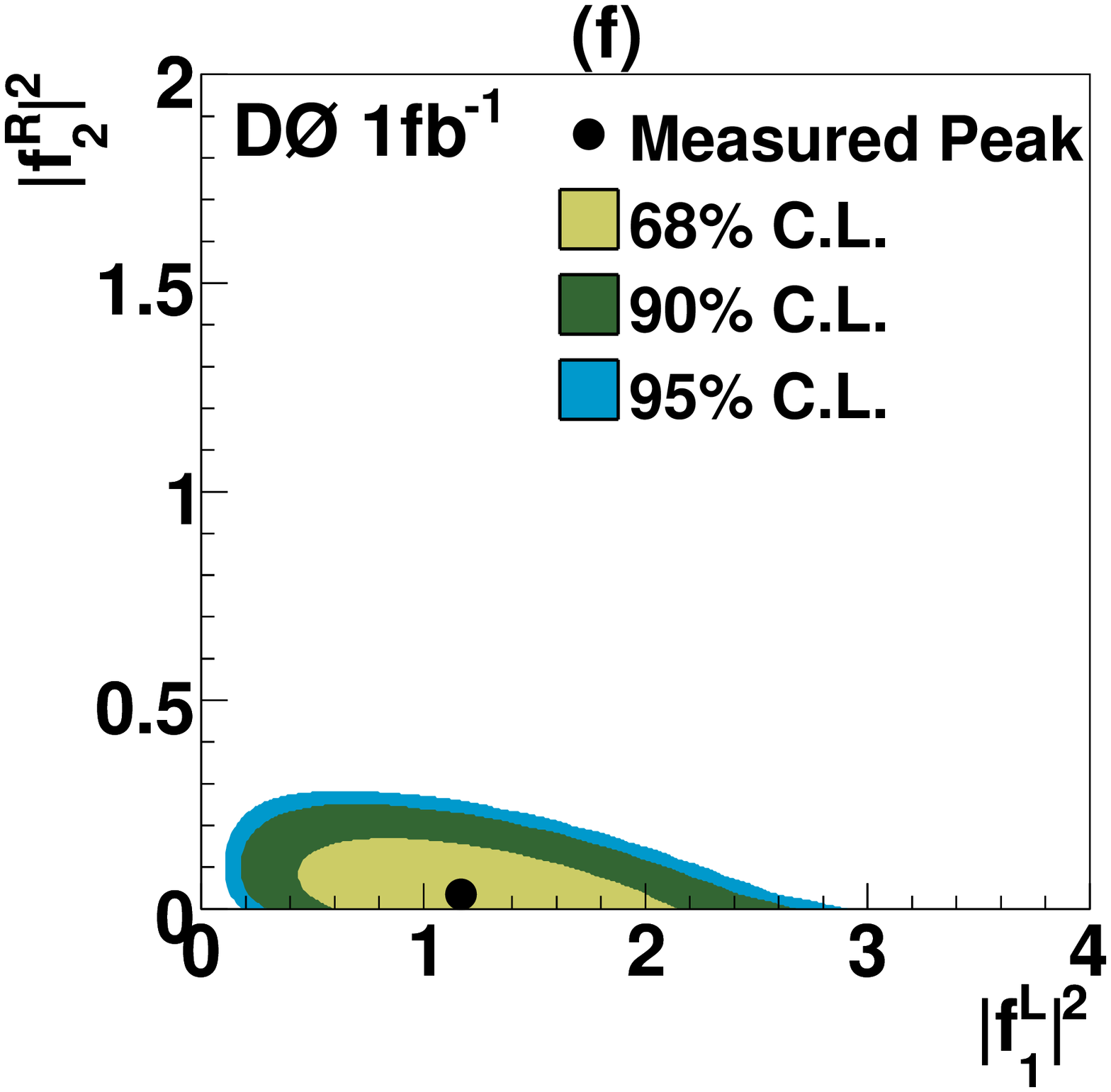}
\vspace{-0.25in}
\caption{$W$~helicity prior (a, c, e) and final posterior density (b, d, f) for right- 
vs left-handed vector coupling (a and b), left-handed tensor vs left-handed vector coupling (c and d), 
and right-handed tensor vs left-handed vector coupling (e and f). The $W$~helicity prior is normalized 
to a peak value of one and shown as equally spaced contours between zero and one. 
The posterior density is shown as contours of equal probability density.
}
\vspace{-0.2in}
\label{fig:measfullsys_2D}
\end{figure}

A binned maximum likelihood fit compares the $\cos\theta^*$ distribution of the selected events to the 
expectations for each $W$ boson helicity state plus background.
We vary both the longitudinal and right-handed helicity fractions $f_0$ and $f_+$ in the fit and find
the relative likelihood of any set of helicity fractions being consistent with the data.  
In the 
previous $W$~helicity publication, we expressed the likelihood in terms of the helicity fractions
and used a prior that was flat in $f_0$ and $f_+$~\cite{Whel-d0}. Here, we instead
express these relative likelihoods in terms of the anomalous $Wtb$ coupling form factors squared
using the relationships given in Ref.~\cite{Chen:2005vr}. The resulting likelihood distributions 
are shown in the left column of Fig.~\ref{fig:measfullsys_2D}.
They show that the $W$~helicity measurement only constrains ratios of the coupling form factors.

We can constrain both the ratios and the magnitudes of the form factors in the single top analysis.
The dominant modes for single top quark production at the Tevatron are the $s$-channel production and
decay of a virtual $W$~boson and the $t$-channel exchange of a $W$~boson. Evidence for production
of single top quarks has been reported by the D0 and CDF collaborations~\cite{abazov:181802,Aaltonen:2008sy}.
Both the cross section and the angular correlations of the final state objects are modified in the presence
of anomalous couplings. The total cross section for SM single top quark production at a top quark
mass of 172.5~GeV is predicted to be  $3.15\pm0.3$~pb~\cite{singletop-xsec-sullivan}. 
For this analysis, we assume that single top quarks are produced exclusively through 
$W$~boson exchange and that the $Wtb$ vertex dominates top quark production and decay.

We look for single top quark production in events with one lepton [electron ($p_T>15$~GeV)
or muon ($p_T>18$~GeV)] and $\met>15$~GeV. We select a sample that is statistically independent of the 
$W$~helicity analysis by asking for two or three jets with $p_T>15$~GeV, of which one should have $p_T>25$~GeV.
We also require at least one of the jets to be identified as originating from 
a $b$~hadron by a $b$-tagging algorithm. Details of the selection criteria and background 
modeling are given in Ref.~\cite{abazov:181802}.

We model the single top quark signal using the {\sc comphep-singletop} MC event 
generator~\cite{singletop-mcgen} where anomalous $Wtb$ couplings are considered in both 
the production and decay of the top quark. The background modeling for the single top analysis
utilizes the same samples as the $W$~helicity analysis for $W$+jets and multijet
backgrounds. The {\ttbar} background in the
single top quark sample is small and is modeled by simulated SM {\ttbar} events. It is 
normalized to the theoretical cross section~\cite{ttbar-xsec-1}.

%
The selection efficiencies for single top quark signals with different $Wtb$ couplings are approximately 
(1--2)\% for events with one $b$~tag and less than 1\% for events with  two $b$~tags. We 
select 1152 events, which we expect to contain $ 56 \pm 12$ SM single top quark 
events. 
We use boosted decision trees~\cite{decisiontrees-breiman,boosting-freund} to extract single
top quark events from the large background.

Systematic uncertainties in the signal and background models are 
described in detail in Refs.~\cite{Whel-d0} and~\cite{abazov:181802}. We take all systematic 
uncertainties and their correlations into account.  
%
Systematic uncertainties in the $W$ boson helicity measurement arise from finite MC 
statistics and uncertainties on the top quark mass, jet energy calibration, and MC models of 
signal and background.  Variations in these parameters can change the measurement in 
two ways: by altering the estimate of the background in the final sample 
(i.e., if the final selection efficiency changes) and by modifying the shape 
of the \coss ~templates. 
Systematic uncertainties in the single top analysis arise from the $W$+jets normalization, the
$W$+jets flavor composition estimate, and the top quark pair background modeling.

Most of the systematic uncertainties are taken to be 100\% correlated between the two
analyses. Systematic uncertainties that affect only the $W$~helicity analysis are MC statistics 
and MC background model. Systematic uncertainties arising from the luminosity
measurement affect only the single top analysis.

We use a Bayesian statistical analysis~\cite{bayes-limits} to combine the $W$~helicity result with
the single top anomalous coupling result. The likelihood result from the $W$~helicity analysis is 
used as a prior to the single top anomalous coupling analysis. 

For any pair of values of the two coupling form factors under consideration, we compare the boosted
decision tree output for the data with the sum of backgrounds and the two signals.
In the scenario where $f_1^L$ and $f_2^L$ are non-zero, the two amplitudes interfere, which we take
into account by using a superposition of three signal samples: 
one with only left-handed vector couplings; one with only 
left-handed tensor couplings; and one with both coupling form factors set to 
one, containing the interference term. We then compute a likelihood as a product 
over all separate analysis channels. We assume Poisson distributions for the observed counts 
and use multivariate Gaussian distributions to model the uncertainties on the combined signal acceptance 
and background yields, including correlations.
The uncertainties are evaluated through MC integration. We generate an ensemble of 5000 samples,
each with a different shift in the various systematic uncertainties, and
compute the Bayesian posterior for each sample. The final posterior is then the ensemble average of 
all individual posteriors. 

The two-dimensional posterior probability density is computed as a function of $|f^{L}_{1}|^2$ and 
$|f_X|^2$, where $f_X$ is $f^{R}_{1}$, $f^{L}_{2}$, or $f^{R}_{2}$. 
These probability distributions are shown in~Fig.~\ref{fig:measfullsys_2D}. In all three 
scenarios we measure approximately zero for the anomalous coupling form factors and favor the 
left-handed vector hypothesis over the alternative hypothesis.
We compute 95\%~Confidence Level (C.L.) upper limits on these form factors by integrating out the left-handed vector 
coupling form factor to get a one-dimensional posterior probability density. The measured values are 
given in Table~\ref{table:obslim}. 

In comparison, the limits at 95\%~C.L. without the 
$W$~helicity constraints are $|f_1^R|^2< 1.83$, $|f_2^L|^2< 0.52$, and $|f_2^R|^2< 0.24$. 
The kinematic distributions of the $f_1^L$ and $f_1^R$ single top quark samples are similar enough that
the single top anomalous coupling analysis in this scenario is mainly sensitive to the total cross section. 
Hence, the $W$~helicity analysis improves the $|f_1^R|^2$ limit significantly. 
Conversely, it does not add much information to the right-tensor coupling limit where most of the
sensitivity is provided by the single top anomalous coupling analysis. 

In summary, we have presented the first study of $Wtb$ couplings that combines $W$~helicity measurements
in top quark decay with anomalous couplings searches in the single top quark final state,
thus using all applicable top quark measurements by D0. We find consistency with the SM and set 95\%~C.L. 
limits on anomalous $Wtb$ couplings. Our limits represent significant improvements over previous results, 
and rule out a right-handed top quark vector coupling form factor of magnitude one for the first time.


\begin{table}[t]
\caption{\label{table:obslim} Measured values with uncertainties and upper limits at the 95\%~C.L. 
for $Wtb$ couplings in three different scenarios.} 
\begin{tabular}{l@{ }l@{ }l} \hline\hline
Scenario  & ~Coupling                   & Coupling limit if $f^L_1=1$    \\
\hline	
\vspace{-0.08in} \\
($L_1,R_1$)    &  $|f^L_1|^2=1.27 ^{+0.57}_{-0.48}$      &   \\
               &  $|f^R_1|^2<0.95 $                      & $|f^R_1|^2< 1.01 $ \\
($L_1,L_2$)    &  $|f^L_1|^2=1.27 ^{+0.60}_{-0.48}$      &   \\
               &  $|f^L_2|^2<0.32 $                      & $|f^L_2|^2< 0.28 $ \\
($L_1,R_2$)    &  $|f^L_1|^2=1.04 ^{+0.55}_{-0.49}$      &   \\
               &  $|f^R_2|^2<0.23 $                      & $|f^R_2|^2< 0.23 $ \\
\hline \hline
\end{tabular}
\vspace{-.1in}
\end{table}

%
We thank the staffs at Fermilab and collaborating institutions, 
and acknowledge support from the 
DOE and NSF (USA);
CEA and CNRS/IN2P3 (France);
FASI, Rosatom and RFBR (Russia);
CNPq, FAPERJ, FAPESP and FUNDUNESP (Brazil);
DAE and DST (India);
Colciencias (Colombia);
CONACyT (Mexico);
KRF and KOSEF (Korea);
CONICET and UBACyT (Argentina);
FOM (The Netherlands);
STFC (United Kingdom);
MSMT and GACR (Czech Republic);
CRC Program, CFI, NSERC and WestGrid Project (Canada);
BMBF and DFG (Germany);
SFI (Ireland);
The Swedish Research Council (Sweden);
CAS and CNSF (China);
and the
Alexander von Humboldt Foundation (Germany).
%
\vspace{-0.1in}



\begin{thebibliography}{99}
\vspace{-0.1in}

%
\bibitem[a]{alton}
Visitor from Augustana College, Sioux Falls, SD, USA.
\bibitem[b]{askew,gershtein}
Visitor from Rutgers University, Piscataway, NJ, USA.
\bibitem[c]{burdin}
Visitor from The University of Liverpool, Liverpool, UK.
\bibitem[d]{hensel,meyer,park,quadt}
Visitor from II. Physikalisches Institut, Georg-August-University,
  G{\"o}ttingen, Germany.
\bibitem[e]{luna-garcia}
Visitor from Centro de Investigacion en Computacion - IPN,
  Mexico City, Mexico.
\bibitem[f]{podesta-lerma}
Visitor from ECFM, Universidad Autonoma de Sinaloa, Culiac\'an, Mexico.
\bibitem[g]{voutilainen}
Visitor from Helsinki Institute of Physics, Helsinki, Finland.
\bibitem[h]{weber}
Visitor from Universit{\"a}t Bern, Bern, Switzerland.
\bibitem[i]{wenger}
Visitor from Universit{\"a}t Z{\"u}rich, Z{\"u}rich, Switzerland.
\bibitem[\ddag]{deceased}
Deceased.

%
\vskip 0.25cm

%
\bibitem{TT_window}
T.~Tait and C.-P. Yuan,  Phys.\ Rev.\ D\ {\bf 63}, 14018 (2001).  
%
\bibitem{cpyuan_0503040v3}
G.L.~Kane, G.A.~Ladinsky, and C.-P.~Yuan, Phys.\ Rev.\ D\ {\bf 45}, 124 (1992).
%
\bibitem{Cabibbo:1963yz}
  N.~Cabibbo,
  Phys.\ Rev.\ Lett.\  {\bf 10}, 531 (1963);
  M.~Kobayashi and T.~Maskawa,
  Prog.\ Theor.\ Phys.\  {\bf 49}, 652 (1973).
%
\bibitem{bsgamma}
F. Larios, M.A. Perez, and C.-P. Yuan, Phys.\ Lett.\ B\ {\bf 457}, 334 (1999);
G. Burdman, M.C. Gonzales-Garcia, and S.F. Novaes, Phys.\ Rev.\ D\ {\bf 61}, 114016 (2000),
and references therein.
%
\bibitem{NIM}
V.M.~Abazov {\sl et al.} (D0 ~Collaboration), Nucl. Instrum. Methods\ A\ {\bf 565}, 
463 (2006). 
%
\bibitem{Whel-d0}
V.M.~Abazov {\sl et al.} (D0 Collaboration), Phys.\ Rev.\ Lett.\ {\bf 100}, 062004 (2008).
%
\bibitem{wtb-prl}
V.M.~Abazov {\sl et al.} (D0 Collaboration), Phys.\ Rev.\ Lett.\ {\bf 101}, 221801
(2008).
%
\bibitem{Chen:2005vr}
  C.R.~Chen, F.~Larios, and C.~P.~Yuan,
  Phys.\ Lett.\ B\ {\bf 631}, 126 (2005).
%
\bibitem{alpgen}
M.L.~Mangano {\sl et al.}, JHEP {\bf 07}, 001 (2003). We used {\sc alpgen} version 2.05.
%
\bibitem{pythia}
T. Sj\"{o}strand {\sl et al.}, arXiv:hep-ph/0308153 (2003). We used {\sc pythia} version 6.323.
%
\bibitem{cteq}
J.~Pumplin {\sl et al.}, JHEP {\bf 07}, 012 (2002); D. Stump {\sl et al.}, JHEP {\bf 10}, 046 (2003).
%
\bibitem{geant}
R.~Brun and F.~Carminati, CERN Program Library Long Writeup W5013, 1993 (unpublished).
%
\bibitem{abazov:181802}
V.M.~Abazov {\sl et al.} (D0 Collaboration), Phys.\ Rev.\ Lett.\ {\bf 98}, 181802 (2007);
V.M.~Abazov {\sl et al.} (D0 Collaboration), Phys.\ Rev.\ D\ {\bf 78}, 12005 (2008).
%
\bibitem{Aaltonen:2008sy}
  T.~Aaltonen {\sl et al.}  (CDF Collaboration), Phys.\ Rev.\ Lett.\ {\bf 101}, 
252001 (2008).
%
\bibitem{singletop-xsec-sullivan}
Z.~Sullivan, Phys.\ Rev.\ D~{\bf 70}, 114012 (2004).
%
\bibitem{singletop-mcgen}
E.~Boos {\sl et al.},
Phys.\ Atom.\ Nucl.\ {\bf 69}, 1317 (2006); %
%
E.~Boos {\sl et al.} (CompHEP Collaboration),
Nucl.\ Instrum.\ Methods Phys.\ Res. A\ {\bf 534}, 250 (2004).
%
\bibitem{ttbar-xsec-1}
N.~Kidonakis and R.~Vogt,  Phys.\ Rev.\ D\ {\bf 68}, 114014 (2003).
%
\bibitem{decisiontrees-breiman}
L.~Breiman {\sl et al.},
{\em Classification and Regression Trees}
(Wadsworth, Stamford, 1984). %
%
\bibitem{boosting-freund}
Y.~Freund and R.E.~Schapire,
in {\em Machine Learning: Proceedings of the Thirteenth International
Conference}, edited by L.~Saitta (Morgan Kaufmann, San Fransisco,
1996) p.~148.
%
\bibitem{bayes-limits}
I.~Bertram {\sl et al.}, FERMILAB-TM-2104 (2000).
%



\end{thebibliography}
\end{document}